\documentstyle[aps,preprint]{revtex}
\input {amssym.def}
\input {amssym.tex}
\everymath{\rm}
\everydisplay{\rm}
\begin{document}
\title{Dynamic Jahn-Teller Effect and  Colossal Magnetoresistance in $
La_{1-x} Sr_xMnO_3$}
\author{A. J. Millis, Boris I. Shraiman and R. Mueller}
\address{AT\&T Bell Laboratories \\
600 Mountain Avenue \\
Murray Hill, NJ 07974}
\maketitle
\begin{abstract}
A model for $La_{1-x}Sr_xMnO_3$ which incorporates
the physics of dynamic
Jahn-Teller and double-exchange effects is presented
and  solved via a dynamical mean field
approximation.
In an intermediate coupling regime
the interplay of these two effects is found to reproduce
the behavior of the resistivity and magnetic transition
temperature observed in
$La_{1-x} Sr_x MnO_3$.
\end{abstract}
\newpage
In this note we present and analyse a
model which captures the important physics of the
``colossal magnetoresistance'' materials
$La_{1-x} A_x MnO_3$
(here A is a divalent element such as Sr or Ca).
In the interesting doping range $0.2 \lesssim  x \lesssim  0.4$,
$La_{1-x} A_x MnO_3$ is a ferromagnetic metal at low
temperature, T, and a poorly conducting paramagnet at high T;
the ferromagnetic-paramagnetic transition occurs at an
x dependent transition temperature $T_c (x) \sim300K$ and is
accompanied by a large drop in the resistivity [1].
The ``colossal magnetoresistance'' which has stimulated the recent
interest in these materials is observed for temperatures near
$T_c (x)$ [2].

Some aspects of the physics of
$La_{1-x} A_x MnO_3$ are well established [1].
The electronically active orbitals are the Mn d-orbitals
and the mean number of $d$ electrons per Mn is
4-x.
The cubic anisotropy and Hund's rule coupling are sufficiently
large that 3 electrons go into tightly bound $d_{xy}$,
$d_{yz}$, $d_{xz}$ core states and make up an electrically inert
core spin $S_c$ of magnitude 3/2; the remaining (1-x) electrons
go into a band of width $\sim2.5$ eV
made mostly of the
outer-shell $d_{x^2 - y^2}$
and $d_{3z^2 -r^2}$ orbitals \cite{Mattheiss95}.
The outer shell electrons are aligned to the core states
by a Hund's Rule coupling $J_H$ which is believed to be
large [1].

The large value of $J_H$ means that the hopping of an outer
shell electron between two Mn sites is affected by the relative alignment
of the core spins, being maximal when the core spins are
parallel and minimum when they are antiparallel.
This phenomenon, called ``double exchange''
\cite{Zener51}, has been widely regarded \cite{Competition,Furukawa94}
as the only significant physics in the regime
$0.2 \lesssim  x \lesssim  0.5$.
However, we have previously shown \cite{Millis95}
that double exchange
alone cannot account for the very large resistivity of the
$T > T_c$ phase \cite{resistivity} or for the sharp drop in resistivity just
below $T_c$.
We suggested that the necessary extra physics is a strong
electron-phonon coupling due at least in part to a Jahn-Teller splitting
of the Mn $d^4$ state in a cubic environment.
The cubic-tetragonal
phase transition observed for $0 \lesssim x \lesssim 0.2$
is known to be due
to a frozen-in Jahn-Teller distortion with long range order
at the wave vector ($\bf{\pi,\pi,\pi}$) \cite{Kanamori59}.
We proposed that for $x > 0.2$ and
$T > T_c (x)$, slowly fluctuating local Jahn-Teller distortions
localize the conduction band electrons as polarons.
The interesting physics issue is then how the polaron effect is
``turned off'' as $T$ is decreased through $T_c$, permitting
the formation of a metallic state.  Our picture is as follows.
The competition between electron itineracy and self-trapping
is controlled by the dimensionless
ratio of the Jahn-Teller self-trapping energy
$E_{J-T}$ and an electron itineracy energy which may
be parametrized by an effective hopping matrix element $t_{eff}$.
When
$E_{J-T} / t_{eff}$ exceeds a critical value we expect a crossover
from a Fermi liquid to a polaron regime.
In models with both double exchange and a large $E_{J-T}$,
an interesting interplay may occur because $t_{eff}$ is affected
by the degree of magnetic order and conversely.
As $T$ is increased from zero, the spins
begin to disorder.
This reduces $t_{eff}$ which increases $E_{J-T} / t_{eff}$ so phonon
effects become stronger, further localizing the electrons
and reducing $t_{eff}$ and thereby the effective ferromagnetic
coupling.

To investigate this quantitatively
we consider the model Hamiltonian $H_{eff} = H_{el} +
H_{J-T}$ with
\begin{eqnarray}
H_{el} &=& - \sum_{ij \alpha}
t_{ij}^{ab} d_{ia \alpha}^\dagger
d_{j b \alpha} +J_H \sum_{i,a, \alpha}
\vec{S}_c^i \cdot
d_{ia \alpha}^\dagger
\vec{\sigma} d_{ia \alpha}+\vec{h} \cdot \vec{S}_c/S_c
\end{eqnarray}
and
\begin{eqnarray}
H_{J-T} &=& g \sum_{ja \sigma} d_{ja \sigma}^\dagger
{\bf Q}^{ab}(j) d_{jb \sigma}
+k \sum_{j} {\bf Q}^2(j).
\end{eqnarray}.

Here  $d_{a \sigma}^\dagger(i)$ creates an outer-shell d-electron of
spin $\sigma$ in the a orbital on site i.  The local lattice distortions
which cause the Jahn-Teller splitting transform
as a two-fold degenerate
representation of the cubic group which we
parametrize by a magnitude $r$
and an angle $\phi$.  They couple to the electron as a traceless
symmetric matrix
${\bf Q} = r(cos(\phi) {\bf \tau}_z+sin(\phi){\bf \tau}_x)$.
The electron-phonon coupling is g and the phonon stiffness is k.
The external magnetic field is $\vec{h}$; for simplicity
we have coupled it to the core spin only.
In the phonon part of $H_{eff}$ we have neglected
intersite terms and also cubic and higher nonlinearities.
In the electronic part of $H_{eff}$ we have neglected
on-site Coulomb interaction
effects; these will be important for higher energy properties
of spectral functions but will only affect the low-energy properties
we consider here by renormalizing parameters such as $t^{ab}_{ij}$.

To solve $H_{eff}$ we introduce further simplifications.
We take  $J_H \rightarrow \infty$.
Because we are interested in phenomena at temperatures of order room
temperature, we assume the phonons and the core spins are classical.
We allow
for  magnetic order but assume that there
is no long range order in the lattice degrees of
freedom.  To solve the electronic
problem we use the "dynamical mean field" approximation which becomes
exact in a limit in which the  spatial dimensionality
$d \rightarrow \infty$ \cite{Kotliar95}.
Then, the free energy may be expressed in
terms of a space-independent "effective field" ${\bf G_{eff}}(\omega)$
via
\begin{equation}
Z=\int r dr d\phi d\Omega exp[-t r^2/2 T+
Tr ln[t{\bf G_{eff}}^{-1}
+\lambda \vec{r} \cdot \vec{{\bf \tau}}+
J_H \vec{S_c}\cdot \vec{{\bf \sigma}}]+\vec{h} \cdot \vec{\Omega}]
\end{equation}
Here $\vec{\Omega}$ is the direction of $\vec{S}_c$
and $t=D/4$ (D is the full bandwidth, so from \cite{Mattheiss95}
one estimates $t \approx 0.6eV$).
The dimensionless
electron-phonon coupling constant $\lambda = g/\sqrt{kt}$.
${\bf G_{eff}}(\omega)$  is
a tensor with orbital and spin indices; it
obeys a self-consistency condition whose  form
depends upon the lattice
whose $d \rightarrow \infty$ limit is taken \cite{Kotliar95}.
We have used the Bethe lattice equation, which corresponds
to an underlying band structure with a semicircular density of
states with  $D=2Tr{\bf t}^2$.
The self consistent equation is \cite{Kotliar95}
\begin{equation}
{\bf G_{eff}}^{-1}(\omega) = \omega-\mu-Tr[{\bf tGt}]/2
\end{equation}
where ${\bf G}=\partial ln Z/\partial {\bf G_{eff}}^{-1}$.
Because we assume there is no long range order in the lattice degrees
of freedom, we take ${\bf G_{eff}}$ to be the unit matrix in orbital space.
We have used two methods for treating the spin part of the problem.
In the ${\it direct}$ ${\it integration}$ method,
one  solves Eq. (4)
by performing the integrals over the angle and phonon
coordinates numerically. In the ${\it projection}$ ${\it method}$,
one quantizes the electron spin on site i along an
axis parallel to $\vec{S}_c^i$  and
retains only the component parallel to $\vec{S}_c^i$.
The $J_H$ term then drops out of the Hamiltonian but as shown
previously \cite{Zener51,Millis95},
one must multiply $t_{ij}$
by the double exchange factor $q_{ij}= cos(\theta_{ij}/2) =
\sqrt{(1+\vec{S}_c^i \cdot \vec{S}_c^j)/2}$.  Within
mean field theory one may replace $q_{ij}$ by $q=(1+m^2/2)/\sqrt{2}$,
where $\vec{m}=\langle \vec{S}_c^i \rangle/S_c$  is determined
self consistently via $m=-T\partial/\partial h[
coth\beta(Jm+h)-(\beta(Jm+h))^{-1})$ and
as  shown previously \cite{Millis95}, $J=(1/2\sqrt{2})
\partial lnZ/\partial t$
with Z evaluated at $q=\sqrt{(1+m^2)/2}$.
 The resulting $d=\infty$ equations involve
a $G_{eff}$ which is a scalar and a numerical integral over the
phonon coordinate only.
Because $t$ enters the mean field equations
only as an energy scale, it is necessary only to solve
the resulting
mean field equations once at each T and $\lambda$ to yield
a $Z(T/t,\lambda^2 /t)$; the
q dependence and hence the magnetic properties may  be
found by scaling $t \rightarrow qt$.
The two approaches give very similar
results  for  the magnetic phase boundary and the phonon contribution
to the resistivity, but the direct integration approach gives also the
spin disorder contribution to the resistivity.
The main difference between the two
calculations is that projection method leads to a
first order magnetic phase transition for $\lambda \approx 1$.

We now discuss  the solutions.
Several soluble limits exist.
At $\lambda=0$ there is a second order
ferromagnetic transition at a
$T_c(x)$ which is maximal for the half filled band ($T_c(0)=0.17t$)
and decreases as the band filling is decreased.
For $T > T_c$ there is spin disorder scattering which (in the mean field
approximation used  here) is temperature independent and of small
magnitude.  This scattering decreases below $T_c$ as
discussed previously \cite{Competition,Furukawa94,Millis95}.

In the limit $T \rightarrow 0$, ground state is a fully polarized
ferromagnet for all $\lambda$.
The phonon probability distribution $P ( r ) = \int
d \phi d \Omega
e^{-t r^2/2 T+Trln[t{\bf G_{eff}}^{-1}+\lambda \vec{\tau}
\cdot \vec{r}]}$
is sharply peaked about the most probable
value $r = r^*$.
For $\lambda < \lambda_c (x)$,
$r^* =0$ and the ground state is a conducting Fermi liquid.
For $ \lambda_{eff} > \lambda_c$, $r_* >0$, implying
a frozen-in lattice
distortion and, if $r_*$ is large enough,
a gap in the electronic spectrum.
For $x=0$, $\lambda_c =1.08...$ and the transition is
second order, with $r_*$ linear in $\sqrt{\lambda-\lambda_c}$ but
very rapidly growing, reaching the point  $r_*=1$  at
$\lambda =1.15$.  For $r_* > 1$, $r_*$ becomes linear in $\lambda$
and a gap appears in the spectral function.
For $x > 0$ the transition is first order, involves a jump to
a nonzero $r_*$, and occurs at a $\lambda_c > 1.08$.  A detailed
discussion of the spectral functions and transitions will be presented
elsewhere \cite{Unpublished}
The increase of $\lambda _c$ with $x$ is due to the
increased kinetic energy per electron.
Note that the double exchange effect
means that the kinetic energy is maximal in the fully polarized
ferromagnetic state.  For uncorrelated spins, the
kinetic energy is smaller by a factor of $\sqrt{2}$
and $\lambda_c$ smaller
by a factor of $2^{1/4}$. In other words,
there is a regime of parameters in which the electron-phonon
interaction is insufficient to localize the electrons at $T=0$
but sufficient to localize them at $T>T_c(x)$.

Another analytically solvable limit is $\lambda \gg 1$.
In this limit an expansion in $1/\lambda$ may be constructed
for arbitrary $1/\lambda T$, and in the leading order one may
evaluate the $r$ integral by steepest descents.
Here we find a second order phase transition
at $T_c= t/12 \lambda ^2$ separating two insulating phases with slightly
different gaps.

We turn now to numerical results,limiting ourselves to  x=0
for simplicity.

The $J_H \rightarrow \infty$ limit means that the
d-bands arising from the outer-shell orbitals are half filled, so the
chemical potential $\mu =0$.  Eq (4) is solved
on the Matsubara axis by direct iteration
starting with the $\lambda =0$ solution.
{}From this solution  Z is constructed and  $\langle m \rangle$
and $\langle r \rangle$ are computed.  The conductivity
is calculated
following \cite{Kotliar95}; the requisite
${\bf G_{eff}}$ on the real axis is
obtained by solving Eq. 4 for real frequencies, using the previously
obtained Matsubara solution to define $Z$.

Fig 1. shows the phase diagram in the $T-\lambda$ plane.
The solid line is a second order  transition
separating ferromagnetic (F) and paramagnetic (P)
regions obtained via the direct solution method.
The light dashed lines separate
regions of weak electron-phonon
coupling in which $d \rho /dT > 0$ from regions of strong electron-phonon
coupling in which $d \rho /dT < 0$.
We identify these regions as metal (M) and insulator (I) respectively.
The $T$ dependence of
the $d\rho/dT$ line below $T_c$ is due mostly to
the temperature dependence of the magnetization.
Increasing
$\lambda$ decreases $T_c$; the variation is particularly
rapid in the crossover region $\lambda \sim 1$,
consistent with the very rapid dependence of $r_*$ on $\lambda$
mentyioned above.  Here also
the magnetic transition calculated via direct integration
becomes more nearly first order.
The projection method
leads to a region of two-phase coexistence for $0.92 <\lambda <1.1$.
This is shown on fig 1 as the area between the heavy dotted line and
the solid line. The different behavior of the two models suggests
that in the crossover region the order of the transition is sensitive
to the approximations of the model.
Other physics, not included here, will also tend to drive
the transition first order.  We mention in particular
anharmonicity in the elastic theory, which
will couple the Jahn-Teller distortions to the uniform strain, and
also the conduction electron contribution
to the binding energy of the
crystal, which produces the observed volume change at $T_c$
\cite{Hwang95}.

The inset to fig 1 shows the average of the square of the
lattice displacement.  This would be measurable in a
scattering experiment sensitive to ${\it rms}$ oxygen
displacements.  In the classical model used here, $r^2 \rightarrow
r_*^2 +T$ as $T \rightarrow 0$.  One sees
that for intermediate couplings the
high temperature
state has a non-zero extrapolation to $T=0$ while the low T state
has a vanishing extrapolation, while for larger couplings both
sides of the transition have non-zero but  different extrapolations.

Fig 2 shows the temperature dependence of the
calculated resistivity for several different values
of $\lambda$.  At  small $\lambda$ and $T> T_c$, $\rho$ is
small and has a $T$-independent piece due to the spin disorder
and a T-linear piece (difficult to
percieve on the logarithmic scale used in fig 2),
due to electron-phonon scattering. As $T$ is
decreased through $T_c$, $\rho$ drops as the spin scattering
is frozen out and the phonon contribution changes slightly.
For larger
$\lambda$ a gap opens in the electron spectral function at $T>T_c$
and the resistivity rises as T is lowered to $T_c$ and then drops
sharply below $T_c$, as the gap closes and metallic behavior
is restored.  Finally, at still
stronger coupling, insulating behavior occurs on both sides
of the transition, although there is still a pronounced drop
in $\rho$ at $T_c$.

Fig 3 shows the magnetic field dependence of $\rho$ for $\lambda=1.12$,
demonstrating that in this region of the phase diagram the "colossal
magnetoresistance" phenomenon occurs.  The magnetic field scale is
too large relative to experiment (as is the calculated
$T_c$), but is very small in comparison to the microscopic
scales of the theory.

The series of resistivity curves presented in fig 2 bears a striking
resemblance to measured resistivities on the series
$La_{1-x}A_xMnO_3$.  We have already noted that
the calculation, which neglects long range order, gives the
generic behavior at any carrier concentration.
We identify the experimental doping x with
the relative strength of the electron phonon interaction, $\lambda$
because increasing x
increases the kinetic energy per electron.
With this identification the
results are consistent with the observed
variation of $T_c$ and $\rho$ with x, and also with the opening
of a gap observed \cite{Okimoto95} in the optical conductivity.
Note also that
as T moves  through $T_c$, the effective ferromagnetic $J$
should drop by about 10 percent,
producing a perhaps observable shift
in the position of the zone boundary magnon.

The present theory is consistent with a recent study of
a $La ( Pr, Y ) Ca_{.3} MnO_3$
series of compounds \cite{Hwang95}.  The substitution
of $Pr, Y$ for $La$
decreases the effective
d-d overlap, decreasing $t$ and increasing $\lambda$.
Experimentally, it
results in a shift of $T_c$ to lower temperatures and
an increasing resistivity anomaly, as found in the calculation.
The observed first
order transition also occurs in one of the mean field theories
we have considered.

In summary, we have presented a solution of a model
describing the Jahn-Teller and double exchange physics, and have
shown that it
accounts naturally for
the existence of a high-$T$ insulating phase, the dramatic
changes
of resistivity at $T_c$, and the extreme sensitivity
to magnetic field.
In obtaining this behavior the interplay of
polaron and double exchange physics is essential.
We have solved the $d = \infty$ equation appropriate to Eq. 3 with $J_H=0$
and found metallic behavior (with $\rho$ and $r^2$ linear in T)
at temperatures greater than the $T=0$ gap  and insulating behavior at
temperatures
less than the $T=0$ gap.
A more detailed exploration of the MF theory allowing uniform or
staggered
ordering of the lattice distortions and varying carrier concentration
will be needed to study
the structural transition at low x and the "charge ordered"
phase \cite{Tokura94,Chen95} at $x \approx .5$.
These calculations however must also
include the intersite phonon coupling.

We acknowledge stimulating discussions with G.Aeppli,
S-W. Cheong, A. Georges, H. Hwang, B. G. Kotliar, H. Monien,
A. Ramirez, T. M. Rice, M. Rozenberg, P. Schiffer and R. Walstedt.
We are particularly grateful to P. B. Littlewood,
who stimulated our interest in the problem, collaborated in the
early stages of this work, and has been a continuing source
of help and encouragement.
A. J. M.  thanks the Institute Giamarchi-Garnier
and the Aspen Center for Physics and B. I. S. the Ecole
Normal Superiure for hospitality.  R. M. was supported
in part by the Studienstiftung des Deutschen Volkes.
\newpage
Figure Captions

Fig. 1:  Phase diagram.  Solid line:  ferromagnetic $T_c$ as a function
of electron-phonon coupling, $\lambda$, calculated by direct
integration method.  The area enclosed by the
solid line and
the heavy dashed line is the region of metastability found
in one formulation of mean field theory. Light dotted
lines:  metal-insulator crossover obtained from calculated resistivities.
Regions labelled
as PM (paramagnetic metal), FM (ferromagnetic metal), PI (paramgnetic
insulator) and FI (ferromagnetic metal) according to the value
of the magnetization and $d\rho/dT$.
Inset:  square of average lattice distortion plotted vs temperature for
$\lambda =0.71 (lowest), 0.9, 1.05, 1.12, 1.2$

Fig. 2:  Resistivity calculated by direct integration
method plotted versus temperature for different couplings
$\lambda$.  Heavy solid curve (top) $\lambda = 1.2$, heavy dotted curve,
$\lambda =1.12$, heavy dot-dash curve $\lambda=1.05$, light solid
curve $\lambda= 0.95$, light dashed curve $\lambda = 0.85$, light dot-dashed
curve (bottom) $\lambda =0.71$.

Fig 3.  Magnetic field dependence of resistivity calculated
by direct integration method for $\lambda = 1.12$
and magnetic field h as shown.  Note h=0.01t corresponds to 15 Tesla
if $t = 0.6eV$ and $S_c = 3/2$.


\begin{references}
\bibitem{Furukawa94}
N. Furukawa, J. Phys. Soc. Jpn. {\it 63},
3214 (1994) and unpublished.
This work presents $d=\infty$ calculations
for the model with $\lambda =0 $; it focusses on the
resistance as a function of ${\it magnetization}$
(which agrees well with data)
and does not note
the difficulty with the magnitude of the resistance.
\bibitem{resistivity}
The observed resistivity at $T > T_c$ is of the
order of $10,000 \mu \Omega -cm$, much greater than
the Mott limit \cite{Tokura94}.
Indeed, if the resistivity is modelled in terms
of a density $x$ of classical particles hopping incoherently with
probability $W$, so
$\rho = 3a_0 k_B T/ xe^2 W$
($a_0 \approx 4A$ is the Mn-Mn distance), one
finds $W \sim10^{11} /sec$ so
$ W \sim 10K \ll k_B T$ and a classical picture
is self consistent.
\end{references}
\end{document}